\documentclass[showpacs,preprintnumbers,amsmath,amssymb,preprint]{revtex4}

\usepackage{graphicx}

\usepackage{dcolumn}

\usepackage{bm}

\preprint{Preprint ESI 1647, TUW--05--07}

\begin{document}

\title{Dynamical instability of fluid spheres in the presence of a cosmological constant}

\author{C. G. B\"ohmer}

\email{boehmer@hep.itp.tuwien.ac.at}
\affiliation{The Erwin Schr\"odinger International Institute for Mathematical
             Physics, Boltzmanngasse 9, A-1090 Wien, Austria}
\affiliation{Institut f\"ur Theoretische Physik, Technische Universit\"at Wien,
             Wiedner Hauptstr. 8-10, A-1040 Wien, Austria}

\author{T. Harko}
\email{harko@hkucc.hku.hk}\affiliation{Department of Physics, The
University of Hong Kong, Pokfulam Road, Hong Kong SAR, P. R.
China}

\date{\today}


\begin{abstract}
The equations describing the adiabatic, small radial oscillations
of general relativistic stars are generalized to include the
effects of a cosmological constant. The generalized eigenvalue
equation for the normal modes is used to study the changes in the
stability of the homogeneous sphere induced by the presence of the
cosmological constant. The variation of the critical adiabatic
index as a function of the central pressure is studied numerically
for different trial functions. The presence of a large
cosmological constant significantly increases the value of the
critical adiabatic index. The dynamical stability condition of the
homogeneous star in the Schwarzschild-de Sitter geometry is
obtained and several bounds on the maximum allowable value for a
cosmological constant are derived from stability considerations.
\end{abstract}



\maketitle

\section{Introduction}

The possibility that the cosmological constant be nonzero and
dominates the energy density of the universe today is one of the
most intriguing problems of the contemporary physics. Considered
already in 1896 by von Seeliger and Neumann (see \citet{ScMi94}
and references therein), who added it to the Poisson equation for
the Newtonian potential to compensate the energy density of the
'aether', introduced by \citet{Ei17} in his equations of general
relativity to obtain a static model of the Universe, eliminated by
him after the discovery of the redshift of the stars by Hubble,
the cosmological constant was reintroduced next by \citet{BoGo48}
and \citet{HoNa62} to resolve an age crisis and to construct a
universe that satisfied the ``Perfect Cosmological Principle''.
The cosmological constant appeared again in inflationary
models of the universe introduced by \citet{Gu81}. The density of
the positive isotropic energy density of the scalar field
(inflaton) that dominates the first stages of the evolution of the
Universe behaves like a cosmological constant leading to a rapid
cosmological expansion of the universe during a de Sitter phase
\citep{Li90}.

Within the classical general relativity, the existence of a
cosmological constant is equivalent to the postulate that the
total energy momentum tensor of the Universe $T_{ik}^{(U)}$
possesses an additional piece $T_{ik}^{(V)}$,
besides that of its matter content $T_{ik}^{(m)}$, of the form $%
T_{ik}^{(V)}=\Lambda g_{ik}$, where generally one may assume that the cosmological constant $%
\Lambda $ is a scalar function of space and time coordinates. Such
a form of the additional piece $T_{ik}^{(V)}$ has previously been
obtained in certain field-theoretical models and is interpreted as
a vacuum contribution to the energy momentum tensor
\citep{ZN71,Dr74,Ad82,OT86},
\begin{equation}
      \Lambda g_{ik}=\frac{8\pi G}{c^{4}}\left\langle T_{ik}^{(V)}\right\rangle =%
      \frac{8\pi G}{c^{4}}\left\langle \rho _{V}\right\rangle g_{ik},
\end{equation}
where $\rho _{V}$ is the energy of the vacuum (see \citet{Pa03}
for a recent review of the cosmological constant problem). The
vacuum value of $T_{ik}$ thus appears in the form of a
cosmological constant $\Lambda $ in the gravitational field
equations. In the framework of standard general relativity the
contracted Bianchi identity requires $\Lambda $ to be a constant.
However, generalized physical models with time and space varying
cosmological constant (decaying vacuum energy) have been
intensively investigated in the physical literature \cite{dec}.

A natural thermal interpretation of the cosmological constant has
been proposed based on the fact that the de Sitter vacuum can be
thought as being hot, since even a geodesic observer in this
manifold will detect an isotropic background of thermal radiation
with temperature $T_{V}=\left( \Lambda /12\pi^{2}\right)^{1/2}$
\citep{Ga88}.  The cosmological constant can thus be interpreted
as a parameter measuring the intrinsic temperature of the empty
space-time, or in a sense, of the geometry itself.

The first pressing piece of data, which motivated a
reconsideration of the cosmological constant involved the study of
Type Ia Supernovae. Observations of Type Ia Supernovae with
redshift up to about $z\sim 1$ provided evidence that we may live
in a low mass-density Universe, with the contribution of the
non-relativistic matter (baryonic plus dark) to the total energy
density of the Universe of the order of $\Omega_{m}\sim 0.3$
\citep{Ri98,Pe98,Pe99}. The value of $\Omega_{m}$ is
significantly less than unity \citep{OsSt95}, and consequently
either the Universe is open or there is some additional energy
density $\rho $ sufficient to reach the value $\Omega _{{\rm
total}}=1$, predicted by inflationary theory. Observations also
show that the deceleration parameter of the Universe $q$ is in the
range $-1\leq q<0$, and the present-day Universe undergoes an
accelerated expansionary evolution \citep{Ri04}.

Several physical models have been proposed to give a consistent
physical interpretation to these observational facts. One
candidate, and maybe the most convincing one for the missing
energy is the vacuum energy density or the cosmological constant
$\Lambda $ \citep{Ei17,Pa03}. The presence of a cosmological
constant implies that, for the first time after inflation, in the
present epoch its role in the dynamics of the Universe becomes
dominant \citep{PeRa03}.

On the other hand there is the possibility that scalar fields
present in the early universe could condense to form the so named
boson stars. There are also suggestions that the dark matter could
be made up of bosonic particles. This bosonic matter would
condense through some sort of Jeans instability to form compact
gravitating objects. A boson star can have a mass comparable to
that of a neutron star \citep{Ry97}. A detailed analysis of the
mass dependence of boson stars on the self-interacting potential
done by \citet{MiSc00} has shown that the mass $M$ of the bosons
stars can range from $M=10^{10}$ kg for mini-boson stars with
radius $R\approx 10^{-18}$ m to values that can reach or extend the
limiting mass of $3.23$$M_{\odot }$ of the neutron stars. For the
very light universal axion of the effective string models, the
gravitational mass is in the range of $\sim 0.5$$M_{\odot}$. The
density of a mini-boson star can exceeds by a factor of $10^{45}$
the density of a neutron star.

In general, a boson star is a compact, completely regular
configuration with structured layers due to the anisotropy of
scalar matter, an exponentially decreasing 'halo', a critical mass
inverse proportional to the constituent mass, an effective radius
and a large particle number (for a detailed review of the boson
star properties see \citep{ScMi03}). The simplest kind of boson
star is made up of a self-interacting complex scalar field $\Phi $
describing a state of zero temperature. The self-consistent
coupling of the scalar field to its own gravitational field is via
the Lagrangian \citep{ScMi03},\citep{SeSu90},\citep{Ku91},
\begin{equation}
      L=\frac{1}{2k}\sqrt{-g}R+\frac{1}{2}\sqrt{-g}\left[ g^{ik}
      \Phi_{;i}^{\ast }\Phi _{;k}-V\left( \left| \Phi \right| ^{2}\right)
      \right] ,
\end{equation}
where $V(|\Phi|^{2})$ is the self-interaction potential usually taken
in the form $V(|\Phi|^{2})=m^{2}|\Phi|^{2}+\lambda |\Phi|^{4}/4$.
Here $m$ is the mass of the scalar field particle (the boson) and
$\lambda$ is the self-interaction parameter. For the bosonic
field the stationarity ansatz is assumed,
$\Phi(t,r) =\phi(r)\exp(-i\omega t)$.
If we suppose that in the star's interior
regions and for some field configurations $\Phi$ is a slowly
varying function of $r$, so that it is nearly a constant, then in
the gravitational field equations the scalar field will play the
role of a cosmological constant, which could also describe a
mixture of ordinary matter and bosonic particles.

At interplanetary distances, the effect of the cosmological
constant could be imperceptible. However, \citet{CaTe98} have
shown that a bound of $\Lambda $ can be obtained using the values
observed from the Mercury's perihelion shift in a Schwarzschild-de
Sitter space-time, which gives $\Lambda <10^{-55}{\rm cm}^{-2}$.
For compact general relativistic objects the cosmological constant
modifies the mass-radius ratio $M/R$ of compact general
relativistic objects. Generalized Buchdahl type inequalities for
$M/R$ in the presence of a cosmological constant have been derived
by \citet{Ma00} and by \citet{Bo04a}, see also \citet{St00}. Thus, for a non-zero
$\Lambda$, the mass-radius ratio is given by $2M/R\leq
4/9+(2/3)(4/9-\Lambda R^2/3)^{1/2}$ \citep{Bo04a}. For a recent
discussion of the astrophysical bounds on the cosmological
constant obtained from the study of spherically symmetric bodies
see \citet{Ba04}.

In a very influential paper, Chandrasekhar \cite{Ch64} studied the
infinitesimal adiabatic oscillations of a gaseous sphere in a
general relativistic framework. In particular, from stability
considerations he derived a condition on the radius $R$ of the
stable dense general relativistic star, which must obey the
constraint $R>\left( 2GM/c^{2}\right) K/\left( \gamma -4/3\right)
$, where $\gamma$ is the ratio of the specific heats of the matter
and $K$ is a constant which depends on the equation of state of
the matter. For homogeneous spheres $K=19/42$, while for
polytropic stars $K$ has values in the range $K=0.565$ ($n=1$) and
$K=1.124$ ($n=3$).  The stability against small radial
oscillations of equilibrium configurations of cold,
gravitationally bound states of complex scalar fields with
equilibrium configurations exhibiting a mass profile against
central density similar to that of ordinary neutron stars was
studied by Gleiser and Watkins \cite{GlWa89}. By studying the
behavior of the eigenfrequencies of the perturbations for
different values of the central density $\sigma $ it was shown by
using both analytical and numerical methods that configurations
with central densities greater than a given value $\sigma _c(0)$
are unstable against radial perturbations. The stability of boson
star solutions in a $D$-dimensional, asymptotically anti-de Sitter
space-time in the presence of a cosmological term was discussed by
\citet{AsRa03}. It was found that for $D>3$ the boson star
properties are close to those in four dimensions with a vanishing
cosmological constant. A different behavior was noticed for the
solutions in the three dimensional case.

It is the purpose of the present paper to study the effects of the
possible existence of a cosmological constant on the radial
oscillations of a dense general relativistic star. As a first step
we generalize the pulsation equation to include $\Lambda$. This
equation is used to study the effect of $\Lambda$ on homogeneous
compact astrophysical objects. The values of the critical
adiabatic index $\gamma $ are estimated numerically for different
values of the cosmological constant $\Lambda $. The stability
criterion of general relativity for a homogeneous fluid sphere is
generalized to take into account the presence of the cosmological
constant. Based on the analysis of the solutions of the
gravitational field equations for a constant density sphere some
other simple criteria of stability for compact objects in the
presence of a cosmological constant are obtained.

This paper is organized as follows. In Section 2, using the
Einstein gravitational field equations, we derive the eigenvalue
equation for adiabatic radial pulsations for the case $\Lambda
\neq 0$. In Section 3 we discuss the mass continuity and
Tolman-Oppenheimer-Volkoff (TOV) equations in the presence of a
cosmological constant and the solution for the homogeneous sphere
is obtained. The stability of the homogeneous star is considered
in Section 4. The results are discussed and summarized in Section
5.

\section[]{Adiabatic radial pulsations of relativistic stars in the Schwarzschild-de Sitter geometry}

In the present Section we will develop the general formalism for
the study of perturbations of the exact solutions of
Einstein's field equations for fluid spheres in the presence of a
cosmological constant, by following the approach introduced by
\citet{Ch64}. We shall consider only perturbations that preserve
spherical symmetry. Under these perturbations only radial motions
will ensue. The metric of the corresponding space-time can be
taken in the form
\begin{equation}\label{line}
      ds^{2}=e^{\nu }\left( dx^{0}\right)^{2}-e^{\lambda}dr^{2}
      -r^{2}\left( d\theta ^{2}+\sin ^{2}\theta d\phi ^{2}\right) ,
\end{equation}
where $\nu$ and $\lambda$ are functions of $x^{0}=ct$ and of the
radial coordinate $r$ only.

The energy-momentum tensor for a spherically symmetric space-time is
\begin{equation}
      T_{i}^{k}=\left(\epsilon +p\right) u_{i}u^{k}-p\delta _{i}^{k},
\end{equation}
where $\epsilon$ is the energy density of the matter and $p$ is
the thermodynamic pressure. $u^{i}=dx^{i}/ds$ is the
four-velocity.

The gravitational field equations corresponding to the line
element given by Eq. (\ref{line}) and with the cosmological
constant included \citep{LaLi}, together with the 0th order
equations, corresponding to the static equilibrium case, are
presented in Appendix A .

Writing $\lambda =\lambda _{0}+\delta \lambda $, $\nu =\nu
_{0}+\delta \nu $, $\rho =\rho _{0}+\delta \rho $ and
$p=p_{0}+\delta p$, where the index $0$ refers to the unperturbed
metric and physical quantities, we find that to the first order
the perturbed components of the energy-momentum tensor are given
by $T_{0}^{0}=\epsilon $, $T_{i}^{i}=-p$, $i=1,2,3$ (no
summation), $T_{0}^{1}=\left( \epsilon _{0}+p_{0}\right) v$ and $%
T_{1}^{0}=\left( \epsilon _{0}+p_{0}\right) ve^{\lambda _{0}-\nu
_{0}}$, where $v=dr/dx^{0}$.

By introducing the ``Lagrangian displacement'' $\xi $ defined as
$v=\partial \xi /\partial t$, we obtain the variations of the
metric functions and of the energy density in a form similar to
the $\Lambda =0$ case: $\delta
\lambda =-\left( \lambda _{0}^{\prime }+\nu _{0}^{\prime }\right) \xi $, $%
\left( \delta \nu \right) ^{\prime }=\left[ \delta p/\left(
\epsilon _{0}+p_{0}\right) -\left( \nu _{0}^{\prime }+1/r\right)
\xi \right] \left(
\lambda _{0}+\nu _{0}\right) $ and $\delta \epsilon =-\left( 1/r^{2}\right) %
\left[ r^{2}\left( \epsilon _{0}+p_{0}\right) \xi \right] ^{\prime
}$, respectively.

In order to obtain an expression for $\delta p$ we need to impose
an extra condition on the system. This condition
is the law of conservation of the baryon number density $n$, which can be written as $%
\left( nu^{i}\right) _{;i}=0$. Generally $n$ is a function of both
energy density and pressure, $n=n\left( \epsilon ,p\right) $. By taking $%
n=n_{0}(r)+\delta n\left( x^{0},r\right) $ it follows that the
variation of the pressure can be obtained as
\begin{equation}
\delta p=-\xi \frac{dp_{0}}{dr}-\gamma p_{0}\left[ \frac{\exp
\left( \nu _{0}/2\right)}{r^{2}}\right] \left[ r^{2}\exp \left(
-\nu _{0}/2\right) \xi \right] ^{\prime },
\end{equation}
where $\gamma $ is the adiabatic index defined as \citep{Ch64}
\begin{equation}
\gamma =\left( p\frac{\partial n}{\partial p}\right) ^{-1}\left[
n-\left( \epsilon +p\right) \frac{\partial n}{\partial \epsilon }
\right].
\end{equation}

With the use of the previous results for the perturbed quantities,
and by assuming that all the perturbations have a dependence on
$x^{0}$ of the form $\exp(i\omega x^{0})$, from the linearized
Einstein field equations we obtain the Sturm-Liouville eigenvalue
equation for the eigenmodes in the presence of a cosmological
constant as
\begin{equation}\label{eig}
\frac{d}{dr}\left( \Pi \frac{d\zeta_j}{dr}\right) +\left(
Q+\omega ^{2}_jW\right) \zeta_j=0,\ j=1,2,...,n,
\end{equation}
where we have redefined the ``Lagrangian displacement'' $\xi$ as
$\xi\rightarrow r^{-2}\exp[\nu /2(r)]\zeta$
(corresponding to a Lagrangian displacement of the radial
coordinate of the
form $\delta r(x^{0},r) =r^{-2}\exp[\nu /2(r)] \zeta \exp(i\omega x^{0})$
and we have denoted
\begin{equation}
\Pi =\frac{\gamma p}{r^{2}}e^{[\lambda (r)+3\nu(r)]/2},
\end{equation}
\begin{equation}
Q=\frac{\Pi }{\gamma p}\left[ \frac{p^{\prime 2}}{\epsilon
+p}-\frac{4p^{\prime }}{r}-\left( \frac{8\pi G}{c^{4}}p-\Lambda
\right)(\epsilon +p) e^{\lambda(r)}\right] ,
\end{equation}
\begin{equation}
W=\frac{\epsilon+p}{r^{2}}e^{[3\lambda(r)+\nu(r)]/2}.
\end{equation}

The boundary conditions for $\zeta(r)$ are that
$\zeta(r)/r^{3}$ is finite or zero as $r\rightarrow 0$
and that the Lagrangian variation of the pressure
$\Delta p=-\left(\gamma pe^{\nu (r)/2}/r^{2}\right) d\zeta /dr$
vanishes at the surface of the star.

For all stellar models of physical interest the frequency spectrum
of the normal radial modes is discrete; the $n$-th normal mode has
$n$ nodes between the center and the surface of the star. The
normal mode eigenfunctions are orthogonal with respect to the
weight function $W(r)$: $\int_{0}^{R}W\zeta _{j}\zeta _{k}dr=0$,
if $j\neq k$.

In the presence of a cosmological constant the system of equations
governing infinitesimal adiabatic radial pulsations is given by
\begin{equation}
\frac{d\xi }{dr}=-\frac{1}{r}\left( 3\xi +\frac{\Delta p}{\gamma p}\right) -%
\frac{\xi }{\epsilon +p}\frac{dp}{dr},  \label{gondek1}
\end{equation}
\begin{eqnarray}
\frac{d\Delta p}{dr} &=&\left[ \frac{\omega ^{2}}{c^{2}}e^{\lambda
(r)-\nu (r)}\left( \epsilon +p\right) r-4\frac{dp}{dr}\right] \xi
+  \label{gondek2} \nonumber\\ &&\left[ \frac{r}{\epsilon
+p}\left( \frac{dp}{dr}\right) ^{2}-e^{\lambda
(r)}\left( \frac{8\pi G}{c^{4}}p-\Lambda \right) \left( \epsilon +p\right) r%
\right] \xi +  \nonumber \\
&&\left[ \frac{1}{\epsilon +p}\frac{dp}{dr}-\frac{4\pi
G}{c^{4}}\left( \epsilon +p\right) re^{\lambda (r)}\right] \Delta
p.
\end{eqnarray}

In Eqs. (\ref{gondek1}) and (\ref{gondek2}) $\xi $ is the relative
radial displacement $\xi =\Delta r/r$, where $\Delta r$ is the
radial displacement. $\Delta p$ is the corresponding radial
displacement of the pressure. Eqs. (\ref{gondek1}) and
(\ref{gondek2}) represents the generalization to the case of the
Schwarzschild-de Sitter geometry of the oscillations equations
introduced by \citet{Go97}.

These pulsation equations have to be integrated with the boundary
conditions $\left( \Delta p\right) _{{\rm center}}=-\left( 3\gamma
p\xi \right) _{{\rm center}}$ and $\left( \Delta p\right) _{{\rm
surface}}=0$.

\section{Static fluid spheres with non-zero cosmological constant}

The basic equations describing the equilibrium of a static fluid
sphere in the presence of a cosmological constant are given by
Eqs. (\ref{eq1})--(\ref {eq3}) in Appendix A. These equations have
to be considered together with the equation $\nu _{0}^{\prime
}=-2p_{0}^{\prime }/\left( \epsilon _{0}+p_{0}\right) $, which can
be obtained from the condition $T_{i;k}^{k}=0$, describing the
conservation of the energy momentum tensor.

With the use of the gravitational field equations and of the
conservation equation the Tolman-Oppenheimer-Volkoff (TOV)
equation, describing the hydrostatic equilibrium of a fluid sphere
in the presence of a cosmological constant $\Lambda \neq 0$, can
be immediately obtained in the form \citep{HaMa00,Ma00,Bo04b}
\begin{equation}
p_{0}^{\prime }=-\frac{\left( \epsilon _{0}+p_{0}\right) \left[ \frac{G}{%
c^{2}}M(r)+\left( \frac{4\pi G}{c^{4}}p_{0}-\frac{\Lambda }{3}\right) r^{3}%
\right] }{r^{2}\left[ 1-\frac{2GM(r)}{c^{2}r}-\frac{\Lambda
}{3}r^{2}\right] },  \label{TOV}
\end{equation}
where we have denoted
\begin{equation}
      M(r)=\frac{4\pi}{c^{2}}\int_{0}^{r}\epsilon_{0}r^{2}dr.
      \label{mass}
\end{equation}

For the TOV equation for boson stars see \citet{ScMi03}.

To obtain a dimensionless form of Eq. (\ref{TOV}) and of the mass
continuity equation, $dM/dr=\left( 4\pi /c^{2}\right) \epsilon
_{0}r^{2}$, we introduce a dimensionless independent variable
$\eta $ and the dimensionless functions
$\varepsilon _{0}\left( \eta \right) $, $P_{0}\left( \eta \right) $ and $%
m\left( \eta \right) $ by means of the transformations $r=a\eta $,
$\epsilon _{0}=\epsilon _{c}\varepsilon _{0}\left( \eta \right) $,
$p_{0}=\epsilon _{c}P_{0}\left( \eta \right) $ and $M=M^{\ast
}m\left( \eta \right) $,
respectively. Here $a$ is a scale factor (a characteristic length), $%
\epsilon _{c}$ the central energy density of the star and $M^{\ast
}$ a characteristic mass. With the use of these transformations we
obtain the mass continuity and TOV equations in the following
dimensionless form
\begin{equation}
\frac{dP_{0}\left( \eta \right) }{d\eta }=-\frac{\left[
\varepsilon _{0}\left( \eta \right) +P_{0}\left( \eta \right)
\right] \left[ m(\eta
)+\left( P_{0}\left( \eta \right) -l\right) \eta ^{3}\right] }{\eta ^{2}%
\left[ 1-\frac{2m(\eta )}{\eta }-l\eta ^{2}\right] },  \label{n2}
\end{equation}
\begin{equation}
\frac{dm}{d\eta }=\eta ^{2}\varepsilon _{0}\left( \eta \right) ,
\label{n1}
\end{equation}
where we denoted
\begin{equation}
a^{2}=\frac{c^{4}}{4\pi G\epsilon _{c}},M^{\ast }=\frac{4\pi
\epsilon _{c}a^{3}}{c^{2}},l=\frac{\Lambda c^{4}}{12\pi G\epsilon
_{c}}.
\end{equation}

Eqs. (\ref{TOV}) and (\ref{mass}) must be integrated with the
boundary conditions $p_{0}\left( R\right) =0$ and $M(0)=0$,
respectively, where $R$ is the radius of the fluid sphere,
together with an equation of state of the form
$p_{0}=p_{0}\left(\epsilon_{0}\right)$. For a wide class of
equations of state one can prove uniqueness and existence of these
solutions, see \citep{Re91,Bo04b}.

In the new variables the equation of state becomes $%
P_{0}=P_{0}\left( \varepsilon _{0}\right) $, while the boundary
conditions are given by $m(0)=0$ and $%
P_{0}\left( \eta _{S}\right) =0$, where $\eta _{S}=R/a$ is the
value of the dimensionless radial coordinate $\eta $ at the
surface of the star. If we suppose that $m$ is an increasing
function of $\eta $, while $\varepsilon _{0}$ and $P_{0}$ are
decreasing functions of the same argument, then it follows that
generally $\varepsilon _{0}\in \left[ 1,0\right] $ and $P_{0}\in
\left[ P_{c},0\right] $, where $P_{c}=P_{0}(0)=p_{0}(0)/\epsilon_{c}$
is the value of the pressure at the center of the star.

A particularly important fluid sphere configuration is the one
corresponding to the constant density case, $\epsilon =\epsilon
_{c}=$constant. This condition gives $\varepsilon _{0}=1$,
$\forall \eta \in \left[ 0,\eta _{S}\right] $, and the mass
continuity equation can be integrated immediately to give
\begin{equation}
m(\eta )=\frac{\eta ^{3}}{3}.
\end{equation}

The solution of the TOV equation with the boundary condition
$P_{0}\left( 0\right) =P_{c}$ is given by
\begin{equation}
P_{0}\left( \eta \right) =\frac{\left(1/3-l\right) \alpha -\sqrt{%
1-\left( 2/3+l\right) \eta ^{2}}}{\sqrt{1-\left(2/3%
+l\right) \eta ^{2}}-\alpha },
\end{equation}
where we denoted
\begin{equation}
\alpha =\frac{1+P_{c}}{1/3-l+P_{c}}.
\end{equation}

The radius $R$ of the star can be obtained from the condition
$P_{0}\left( \eta _{S}\right) =0$ and is given by
\begin{equation}
R=a\frac{\sqrt{6P_{c}\left( 2P_{c}+1\right) -9P_{c}\left( P_{c}+2\right) l}}{%
3P_{c}+1-3l}.
\end{equation}

The values of the central pressure $P_{c}$ depend on the
physically allowed upper limit for the pressure. If we consider
the classical restriction of general relativity, $p\leq \epsilon /3$, as $p\geq 0$, it follows that $%
P_{c}\in \left[ 0,1/3\right] $. A more general restriction can be
obtained by assuming that $p\leq \epsilon $, the upper limit being
the strong equation of state for the hot nucleonic gas. It is
believed that matter actually behaves in this manner at densities
above about ten times the nuclear density, that is, at densities
greater than $10^{17}$ g/cm$^{3}$ and at temperatures $T=\left( \epsilon /\sigma \right) ^{1/4}>10^{13}$ K, where $%
\sigma $ is the radiation constant \citep{ZN71}. In this case
$P_{c}\in \left[ 0,1\right] $. Hence generally $P_{c}\leq w$,
where $w\in \left[ 0,1\right] $. Consequently, $\alpha \in \left[
3/(1-3l),6/(4-3l)\right] $.

The metric coefficient $\exp \left( -\lambda _{0}\right) $ is given by
\begin{equation}
\exp \left[ -\lambda _{0}\left( \eta \right) \right] =1-\left(
2/3+l\right) \eta ^{2} .
\end{equation}

 From the Bianchi identity and the
matching across the vacuum boundary of the star it follows that
$\exp \left[ \nu \left( \eta \right) \right] =\left[ 1-\left(
2/3+l\right) \eta _{S}^{2}\right] /\left[ 1+P_{0}\left( \eta
\right) \right] ^{2}$, or, equivalently,
\begin{equation}
e^{\nu _{0}\left( \eta \right) }=\frac{\left( 1/3-l\right) ^{2}}{%
\left(2/3+l\right) ^{2}}\left[ e^{-\frac{\lambda _{0}\left( \eta
\right) }{2}}-\alpha \right] ^{2}.
\end{equation}

\section{Stability of homogeneous stars in the presence of
a cosmological constant}

The eigenvalue problem formulated in Eq. (\ref{eig}) can be
re-expressed in a well-known variational form: the extremal values
of the quantity
\begin{equation}
\omega ^{2}=\frac{\int_{0}^{R}\left[ \Pi \left( \frac{d\zeta
}{dr}\right) ^{2}-Q\zeta ^{2}\right] dr}{\int_{0}^{R}W\zeta
^{2}dr},  \label{var}
\end{equation}
are the eigenvalues $\omega _{j}^{2}$ of Eq. (\ref{eig}) and the functions $%
\zeta (r)$ which give the extremal values are the corresponding
eigenfunctions. A sufficient condition for the dynamical
instability of a mass is that the right hand side of Eq.
(\ref{var}) vanishes for some chosen trial function $\zeta $ which
satisfies the boundary conditions \citep{Ch64}, \cite{Ba66}.

In terms of the system of variables introduced in the previous Section Eq. (%
\ref{var}) takes the form
\begin{eqnarray}\label{osc}
a^{2}\omega ^{2}\int_{0}^{\eta _{S}}\frac{\left( 1+P_{0}\right) }{\eta ^{2}}%
e^{\frac{3\lambda +\nu }{2}}\zeta ^{2}d\eta  =\int_{0}^{\eta
_{S}}\gamma P_{0}\frac{e^{\frac{\lambda +3\nu }{2}}}{\eta
^{2}}\zeta ^{\prime 2}d\eta - \nonumber \\
\int_{0}^{\eta _{S}}\frac{e^{\frac{\lambda +3\nu }{2}}}{\eta
^{2}}\left[ \frac{P_{0}^{\prime 2}}{1+P_{0}}-\frac{4P_{0}^{\prime
}}{\eta }-2\left( P_{0}-\frac{3}{2}l\right) \left( 1+P_{0}\right)
e^{\lambda }\right] \zeta ^{2}d\eta .
\end{eqnarray}

Setting $\omega ^{2}=0$ and assuming that the adiabatic index is
constant throughout the star, we obtain the value of the critical
adiabatic index $\gamma _{c}$ as
\begin{equation}
\gamma _{c}=\frac{\int_{0}^{\eta _{S}}\frac{e^{\frac{\lambda +3\nu }{2}}}{%
\eta ^{2}}\left[ \frac{P_{0}^{\prime 2}}{1+P_{0}}-\frac{4P_{0}^{\prime }}{%
\eta }-2\left( P_{0}-\frac{3}{2}l\right) \left( 1+P_{0}\right) e^{\lambda }%
\right] \zeta ^{2}d\eta }{\int_{0}^{\eta
_{S}}P_{0}\frac{e^{\frac{\lambda +3\nu }{2}}}{\eta ^{2}}\zeta
^{\prime 2}d\eta }.
\end{equation}

For $\gamma <\gamma _c$ dynamical instability occurs and the star
collapses. For a homogeneous star in the presence of a
cosmological constant the equilibrium metric coefficients and the
pressure distribution have been obtained in the previous Section.
In order to find the range of values of $\gamma _c$ we need to
chose some explicit functional forms for $\zeta $. For the trial
function $\zeta =\eta ^3$ the variation of $\gamma _c$ as a
function of the central pressure $P_c$ is represented, for
different values of $l$, in Fig. 1.

\vspace{0.2in}
\begin{figure}[h]
\includegraphics{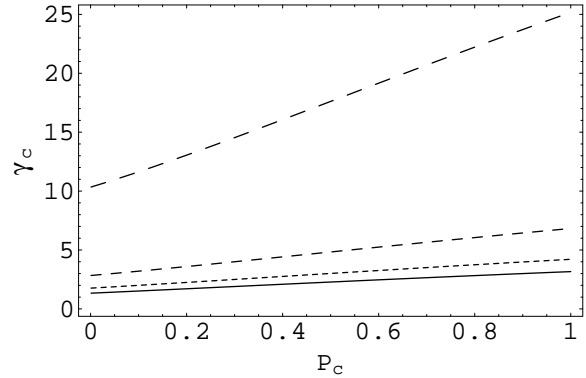}
\caption{ Critical adiabatic index $\gamma _c$ as a function of
the central pressure $P_c$ for the trial function $\zeta =\eta ^3$
and for different values of the cosmological constant: $l=0$
(solid curve), $l=0.1$ (dotted curve), $l=0.2$ (dashed curve) and
$l=0.3$ (long dashed curve).} \label{FIG1}
\end{figure}

In terms of the gravitational radius of the star
$R_{g}=2GM_{S}/c^{2}$, where $M_{S}=4\pi \epsilon _{c}R^{3}/c^{2}$
is the total mass of the star, and of the radius $R$, the critical
adiabatic index $\gamma _{c}$ can be represented as a power series
in $R_g/R$ of the form
\begin{eqnarray}
      \label{gc}
      \gamma_{c} &=&\frac{\frac{4}{3}-l}{1-3l}+
      \frac{19}{42}\left( 1-\frac{21}{19}l\right)
      \left( \frac{R_{g}}{R}\right)\nonumber\\
      &&+\frac{146}{441}\left[ 1-\frac{87}{73}l\left( 1+\frac{21}{58}l\right)
      \right] \left( \frac{R_{g}}{R}\right)^{2} \nonumber \\
      &&+O\left[ \left( \frac{R_{g}}{R}\right)^{3}\right] .
\end{eqnarray}
For the trial function $\zeta =\eta ^3\exp(-\nu /4)$ Fig. 2
represents $\gamma _c$.
\vspace{0.2in}
\begin{figure}[h]
\includegraphics{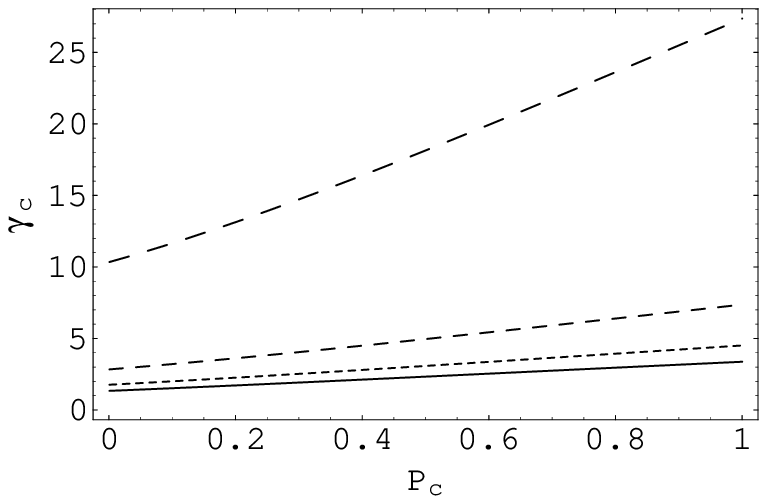}
\caption{ Critical adiabatic index $\gamma _c$ as a function of
the central pressure $P_c$ for the trial function $\zeta =\eta
^3\exp(-\nu /4)$ and for different values of the cosmological
constant: $l=0$ (solid curve), $l=0.1$ (dotted curve), $l=0.2$
(dashed curve) and $l=0.3$ (long dashed curve).} \label{FIG2}
\end{figure}

In this case the representation of $\gamma _c$ in terms of $R_g/R$
is given by
\begin{eqnarray}
      \label{gc1}
      \gamma_{c} &=&\frac{\frac{4}{3}-l}{1-3l}+
      \frac{19}{42}\left( 1-\frac{21}{19}l\right)
      \left( \frac{R_{g}}{R}\right)\nonumber\\
      &&+\frac{5401}{15876}\left[ 1-\frac{26691}{21604}l
      \left( 1+\frac{447}{1271}l\right) \right]
      \left( \frac{R_{g}}{R}\right)^{2}\nonumber\\
      &&+O\left[ \left( \frac{R_{g}}{R}\right)^{3}\right] .
\end{eqnarray}

In first order in $R_g/R$ both Eqs. (\ref{gc}) and (\ref{gc1})
give the same result. This shows that the Chandrasekhar
 stability limit \cite{Ch64} is generally independent of the form
of the trial function.

Lastly, $\gamma_c$ as a function of both $l$ and $P_c$ and for the
trial function $\zeta =\eta^3$, is represented in Fig. 3.
\vspace{0.2in}
\begin{figure}[h]
\includegraphics{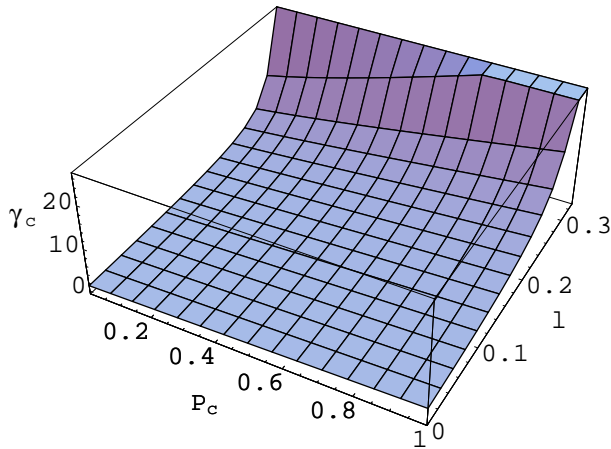}
\caption{ Critical adiabatic index $\gamma _c$ as a function of
the central pressure $P_c$ and of $l$ for the trial function
$\zeta =\eta ^3$} \label{FIG3}
\end{figure}

For a homogeneous sphere and for the trial function $\zeta =\eta
^3$ all the integrals in Eq. (\ref{osc}) can be calculated
exactly. The corresponding pulsation equation and the details of
the calculations are presented in Appendix B.

Performing a series expansion of Eq. (\ref{final}) in Appendix B,
keeping only the terms up to the second order in $\theta_{S}$ and
neglecting the terms containing $l^2$ we obtain the equation
\begin{equation}\label{fin3}
\gamma -\frac{4}{3}-\left( 3\gamma -1\right) l=\frac{390\gamma
-406-3\left( 142+303\gamma \right) l}{378\left(
\frac{2}{3}+l\right) }\theta _{S}^{2}.
\end{equation}

Taking $l=0$ gives
\begin{equation}
\gamma -\frac{4}{3}=\frac{390\gamma -406}{378}\eta _{S}^{2}.
\end{equation}

Therefore
\begin{equation}
      \gamma -\frac{4}{3}<\frac{19}{42}
      \times \frac{2GM_{S}}{c^{2}R},
\end{equation}
or, equivalently,
\begin{equation}
      \label{l}
      R<\frac{19}{42\left(\gamma-4/3\right)}R_{g}.
\end{equation}

Hence from Eq. (\ref{fin3}) we have recovered the stability
condition derived by \citet{Ch64} for $\Lambda =0$.
For $l\neq 0$ we obtain
\begin{equation}
      R<\frac{19}{42}\frac{1-273l/19}
      {\gamma -4\left(1+9l/4\right)/3}R_{g}.
      \label{gen}
\end{equation}

To obtain Eq. (\ref{gen}) we have approximated $\gamma $ in all
terms containing the product $\gamma l$ by $\gamma =4/3$. Eq.
(\ref{gen}) represents the condition for the dynamical stability
of a massive general relativistic object in the presence of a
cosmological constant. It generalizes to the case $\Lambda \neq 0$
the relation initially derived by \citet{Ch64}. In the limit $l=0$
we obtain again Eq. (\ref{l}).

Alternatively, we may divide Eq. (\ref{osc}) by the integral on
its left-hand side and proceed as described above. Up to terms of the
order $\theta_{S}^2$ and $l$ this yields
\begin{equation}
      a^{2}\omega^{2}=\gamma-\frac{4}{3}-l(3\gamma-1)-
      \frac{54\gamma-53-15(4+9\gamma)l}{63(2/3+l)}\theta_S^2,
\end{equation}
which generalizes the original Eq. (77) of \citet{Ch64}.

Approximating, as before, $\gamma$ in all terms containing $\gamma
l$ by $\gamma=4/3$ and letting $\omega\rightarrow 0$ we arrive at
\begin{equation}\label{gen1}
      R<\frac{19}{42}\frac{1-240l/19}
      {\gamma-4\left(1+9l/4\right)/3}R_{g}.
\end{equation}

For the trial function $\xi =\eta ^{3}\exp \left( -\nu /4\right) $
the integrals in Eq. (\ref{osc}) can also be calculated exactly.
A series expansion in terms of $\theta_{S}$ of the result for
this choice gives
\begin{equation}
      \gamma-\frac{4}{3}-\left( 3\gamma -1\right)l=
      \frac{123\gamma-107-3\left( 65+102\gamma \right)l}
      {189\left(2/3+l\right)}\theta_{S}^{2}.
      \label{two}
\end{equation}

For $l=0$ Eq. (\ref{two}) gives the same limit as obtained for the
case of the trial function $\xi =\eta ^{3}$, represented by Eq.
(\ref{l}). For $l\neq 0$, by approximating again $\gamma $ by
$4/3$ in the terms containing the product $\gamma l$, we obtain
\begin{equation}
      R<\frac{19}{42}\frac{1-201l/19}
      {\gamma -4\left( 1+9l/4\right)/3}R_{g},
\end{equation}
which differs only slightly in the numerical value of the
coefficient of $l$ from the previous result. As was already
noted, the slight differences in the approximations show that
firstly, the stability limits are generally independent of
the trial function and that secondly, they are quite unaffected
by the `location' where the higher order terms are neglected.

By imposing the condition $\gamma <\gamma _{c}$ in Eqs. (\ref{gc})
and (\ref {gc1}) gives an other form of the stability criterion
for high density relativistic stars in the presence of the
cosmological constant
\begin{equation}
      \label{gc2}
      R<\frac{19}{42}\frac{1-21l/19}
      {\gamma-\left(4/3-l\right)\left(1-3l\right)^{-1}}R_{g}.
\end{equation}

The denominator of the latter stability criterion coincides with
those derived before, if terms containing $l^2$ are neglected in a
series expansion.

\section{Discussions and final remarks}

The hypothesis that a non-zero cosmological constant could play an
important role in the structure of compact stellar objects seems
to be difficult to accommodate with the smallness of the present
determined or postulated values of this physical quantity. While a
non-vanishing cosmological constant, if definitely confirmed,
would carry a significant implication for our understanding of the
structure and global dynamics of the Universe as a whole, the ways
in which one can clearly test it in an astrophysical or
astronomical setting are very limited. \citet{Pe84} has shown that
the cosmological constant has little effect on the local dynamics.
The effects on the galactic dynamics of $\Lambda$ can hardly be
observed also with distant clusters, because it appears to be
always cancelled in observable quantities \citep{CoSh98}. But, on
the other hand, the effects of the cosmological constant could be
important for stellar objects in the far past, most of the
cosmological models with decaying vacuum energy predicting a time
variable and decreasing cosmological constant \citep{Wa93}.
Therefore, the interior structure of early stars could have been
strongly influenced by the presence of a cosmological constant.

In the presence of a cosmological constant the radius $R$ of the
homogeneous stellar configuration is related to the gravitational
(Schwarzschild) radius of the star $R_{g}$  by the relation
\begin{equation}
R=\frac{\left( 3P_{c}+1-3l\right) ^{2}}{P_{c}\left[ 2\left(
2P_{c}+1\right) -3\left( P_{c}+2\right) l\right] }R_{g}.
\end{equation}

From the restriction $P_c\leq 1$ on the pressure we find the
following simple stability criterion for a homogeneous star in the
presence of a cosmological constant
\begin{equation}
      R\geq \frac{\left( 3w+1-3l\right)^{2}}{w\left[ 2\left(
      2w+1\right) -3\left( w+2\right) l\right] }R_{g}.
\end{equation}

For $w=1/3$ we have $R\geq 9\left( 2-3l\right) ^{2}R_{g}/\left(
10-21l\right) $, while for $w=1$ we obtain $R\geq \left(
4-3l\right) ^{2}R_{g}/\left[3\left( 2-3l\right)\right] $.

The condition of the non-negativity of the radius of the star,
$R\geq 0$, imposes the constraint
$l=\Lambda c^{4}/12\pi G\epsilon_{c}<2(2w+1)/3(w+2)$
on the cosmological constant, or, equivalently,
\begin{equation}
      \label{limit1}
      \Lambda <\frac{\left( 2w+1\right) }{\left( w+2\right) }\frac{%
      8\pi G\epsilon _{c}}{c^{4}}.
\end{equation}

For a star composed from matter having a density equivalent to the
typical nuclear density, $2\times 10^{14}{\rm g/cm}^3$,  Eq.
(\ref{limit1}) leads to an upper limit of the cosmological
constant equal to
\begin{equation}
      \Lambda <3\times 10^{-13}{\rm cm}^{-2}.
\end{equation}

In the presence of a cosmological constant the dynamical stability
condition is modified according to Eq. (\ref{gen}). The critical
value $\gamma _{c}^{\prime }=4/3$ of the adiabatic index is
replaced by $\gamma _{c}^{\prime }=(4/3)(1+9l/4)=
(1+3\Lambda c^{4}/16\pi G\epsilon _{c})$.
A large value of $l$ can increase significantly the value of
$\gamma _{c}^{\prime }$. This could be the case for mixtures of
the potential energy dominated bosonic matter and usual matter,
when the cosmological constant $\Lambda $ can be interpreted as
the static equilibrium value of the scalar field potential,
$\Lambda =V(|\Phi _{0}|^{2})={\rm const.}$
In this case $l$ could have a large value, even of the order of
unity. The presence of the bosonic component strongly affects the
dynamic stability of the system. The critical values of $\gamma
_{c}$ of configurations with given $\epsilon _{c}$ require that
the radius of the star exceeds a certain lower limit $R$, if the
configuration is to be stable. In the presence of a cosmological
constant instability will occur when $\gamma $ is in excess of
$4(1+9l/4)/3$ by a small amount. Since for a perfect gas the
maximum admissible value for $\gamma $ is $\gamma =5/3$, the
stability condition imposes the restriction
\begin{equation}
\left[
5-4(1+\frac{9l}{4})\right] >0,
\end{equation}
or, equivalently,
\begin{equation}
\Lambda <\frac{4\pi G\epsilon _{c}}{3c^{4}}.
\end{equation}

By using the same restriction on $\gamma $ in Eq. (\ref{gc2}) we
obtain
\begin{equation}
\Lambda <\frac{\pi G\epsilon _{c}}{c^{4}}.
\end{equation}

These conditions, obtained from stability considerations, differs
from the condition given by Eq. (\ref{limit1}) only by factors of
the order of unity.

The properties of boson stars and the influence of the
cosmological term on their properties in arbitrary
$D$-dimensional, asymptotically anti de Sitter space-times have
been recently considered in \cite{AsRa03}. By assuming for the
spherically symmetric bound state of the scalar field a stationary
ansatz $\Phi =\phi (r)\exp \left( -i\sigma t\right) $ the metric
for the boson star can be chosen in the form
$ds^{2}=F(r)dr^{2}+r^{2}d\Omega _{D-2}^{2}-F(r)\exp \left[
-2\delta (r)\right] dt^{2}$, where $d\Omega _{D-2}^{2}=\omega
_{ab}dx^{a}dx^{b}$ is the line element of a unit $\left(
D-2\right) $-dimensional sphere and $F(r)=1-2m(r)/r^{D-3}-2\Lambda
r^{2}/(D-2)(D-1)$. The function $m(r)$ is related to the local
mass-energy density up to some dimension-dependent factor. With
the help of these parameters one can define the particle number
$N=\left\{ 4\pi ^{(D-1)/2}/\Gamma \left[ \left( D-1\right)
/2\right] \right\} \int_{0}^{\infty }\phi ^{2}e^{\delta
}F^{-1}r^{D-2}dr$ and the effective radius of the
boson star as $R=\left\{ 4\pi ^{(D-1)/2}/\Gamma \left[ \left( D-1\right) /2%
\right] N\right\} \int_{0}^{\infty }\phi ^{2}e^{\delta
}F^{-1}r^{D-1}dr$. For $\Lambda =0$, boson stars are characterized
by an exponential decay of the scalar field, for which the mass
term in the potential is responsible. For $\Lambda \neq 0$ the
situation is quite different; for $r\rightarrow \infty $, one
finds that $m(r)\sim M+Ar^{2C+D-1}$, where $A$ and $C$ are
some constants depending on the dimensionality of the space-time $D$ and $%
\Lambda $, whose exact form is given in \cite{AsRa03}, and $M$ is
the ADM mass of the star. Thus, the cosmological constant  implies
a complicated power decay at infinity rather than an exponential
one as found in the asymptotically flat space case. The maximum
mass $M_{\max }$ and the maximum particle number $N_{\max }$ for
the boson star decreases with the value of the cosmological
constant, but the general properties of the solution are the same
as in the $\Lambda =0$ case.

The analysis of the stability of the boson stars follows the
standard approach, by assuming that the scalar field and the
metric tensor components can be perturbed in the usual way, with
the perturbation of the scalar field written as $\delta \phi
=f_{1}(t,r)+i\phi _{0}(r)\dot{g}\left( t,r\right) $. Then the
field equations and the Klein-Gordon equation can be linearized to
a system of two self-adjoint coupled equations, the pulsation
equations for the boson stars. These equations have to be
integrated with the boundary
conditions $r^{2}g^{\prime }\rightarrow 0$ at the origin and $%
f_{1}\rightarrow 0$, $g^{\prime }r^{2C+D}\rightarrow 0$ for
$r\rightarrow \infty $.

By assuming that all perturbations are of the form $\exp \left(
i\chi t\right) $, where $\chi $ is the characteristic frequency to
be determined, one can again formulate the stability problem of
the boson stars in the presence of a cosmological constant as a
variational problem, with the eigenvalues $\chi ^{2}$ forming a
discrete sequence $\chi _{0}^{2}\leq \chi _{1}^{2}\leq ...\leq
\chi _{n}^{2}$. The numerical results show that there is a
critical value of the central density of the star $\phi _{c}(0)$
so that $\chi ^{2}>0$ for central values of the scalar field
smaller than $\phi _{c}(0)$ while other configurations are
unstable.

Due to the close mathematical and even physical analogy between
the two eigenvalue problems for boson and "normal" stars, we
expect that stability conditions of the form given by Eqs.
(\ref{gen}), (\ref{gen1}) and (\ref{gc2}) should also exist in the
case of boson stars. For boson stars the corresponding mass
appearing in the stability conditions is the ADM mass $M$, and the
radius is defined as above. The coefficient $\gamma $ becomes a
scalar field-dependent quantity $\gamma _{BS}$, which could be
given as a function of the the central density of the star $\phi
_{c}(0)$, $\gamma _{BS}=\gamma \left[\phi _{c}(0)\right]$. The
values of the numerical coefficients could be also very different
as compared to normal stars. Therefore for boson stars a
mass-radius relation  of the form $R<[f(l)/(\gamma _{BS}-{\rm const})]M$
is a direct consequence of the variational formulation
of the stability problem and of the boundary conditions, requiring
the vanishing of some quantities at center of the star and
infinity, respectively. But finding exact analytical stability
conditions for boson stars requires further mathematical and
numerical investigations.

\acknowledgments
The work of C.~G.~B.~was supported by the Junior Research Fellowship of The
Erwin Schr\"odinger International Institute for Mathematical Physics.

The work of T.~H.~was supported by a Seed Funding Programme for
Basic Research of the Hong Kong Government.

\appendix
\section{Field equations}

For a spherically symmetric matter distribution the gravitational
field equations are given by
\begin{equation}
      \left( re^{-\lambda }\right) ^{\prime }=1-\frac{8\pi G}{c^{4}}%
      T_{0}^{0}r^{2}-\Lambda r^{2},
\end{equation}
\begin{equation}
      \nu^{\prime }=\frac{e^{\lambda }-1}{r}-\frac{8\pi G}{c^{4}}%
      T_{1}^{1}r-\Lambda r,
\end{equation}
\begin{eqnarray}
      -\frac{1}{2}e^{-\lambda }\left(\nu^{\prime \prime }+\frac{\nu ^{\prime 2}}{%
      2}+\frac{\nu ^{\prime }-\lambda ^{\prime }}{r}-\frac{\nu ^{\prime
      }\lambda^{\prime}}{2}\right)\nonumber\\
      +\frac{1}{2}e^{-\nu }\left( \ddot{\lambda}+\frac{\dot{%
      \lambda}^{2}}{2}-\frac{\dot{\lambda}\dot{\nu}}{2}\right)
      =\frac{8\pi G}{c^{4}}T_{2}^{2}+\Lambda ,
\end{eqnarray}
\begin{equation}
      e^{-\lambda }\frac{\dot{\lambda}}{r}=\frac{8\pi
      G}{c^{4}}T_{0}^{1}.
\end{equation}

The zeroth order (or static equilibrium) equations are
\begin{equation}
      \label{eq1}
      \left( re^{-\lambda _{0}}\right) ^{\prime }=1-\frac{8\pi
      G}{c^{4}}\epsilon _{0}r^{2}-\Lambda r^{2},
\end{equation}
\begin{equation}
      \label{eq2}
      \nu _{0}^{\prime }=\frac{e^{\lambda _{0}}-1}{r}+e^{\lambda _{0}}\left( \frac{%
      8\pi G}{c^{4}}p_{0}-\Lambda \right) r,
\end{equation}
\begin{equation}\label{eq3}
\frac{1}{2}e^{-\lambda _{0}}\left( \nu _{0}^{\prime \prime
}+\frac{\nu
_{0}^{\prime 2}}{2}+\frac{\nu _{0}^{\prime }-\lambda _{0}^{\prime }}{r}-%
\frac{\nu _{0}^{\prime }\lambda _{0}^{\prime }}{2}\right) =\frac{8\pi G}{%
c^{4}}p_{0}-\Lambda ,
\end{equation}
where the index $0$ refers to the unperturbed metric and physical
quantities.

\section{Pulsation equation}

In this Appendix we present the pulsation equation for a
homogeneous sphere and for the trial function $\zeta =\eta ^3$. By
introducing the new variable $\theta
=\arcsin\bigl(\sqrt{2/3+l}\eta\bigr)$, denoting $\theta
_{S}=\arcsin\bigl(\sqrt{2/3+l}\eta _{S}\bigr)$ and taking into
account that $\alpha=\cos\theta_{S}/(1/3-l)$, in the limit $\omega
\rightarrow 0$ we obtain the following equation for $\theta_{S}$:
\[ 360\left[
22-34\gamma +9l\left( 1+14\gamma -9l\gamma +3\left( 2+\gamma
\right) l^{2}\right) \right] \theta _{S}\cos \left( \theta
_{S}\right) +
\]
\[
1080\left[ 2+3l+3\gamma \left( 3l-1\right) \right] \theta _{S}\cos
\left( 3\theta _{S}\right) -
\]
\[
20\left[ 134-63l+594l^{3}+27\left( 3l-1\right) \left(
7-8l+6l^{2}\right) \gamma \right] \sin \left( \theta _{S}\right) +
\]
\[
5\left[ -440+717\gamma +9l\left( -22\left( 2+3l^{2}\right)
+9\left( -31+l\left( 13+l\right) \gamma \right) \right) \right]
\times
\]
\[
\sin \left( 3\theta _{S}\right) -
\]
\[
\left[ -10\left( 3l-4\right) \left( 2+3l\right) ^{2}+27\left(
3l-1\right) \left( 7+\left( 8+3l\right) l\right) \gamma \right]
\times
\]
\begin{equation}\label{final}
\sin \left( 5\theta _{S}\right) =0.
\end{equation}

\label{lastpage}

\end{document}